\def   \ni {\noindent}

\def   \ssk {\vskip  5truept}

\def   \bsk {\vskip 15truept}

\def   \newline {\hfil\break}

\def \ref{\noindent}

\documentstyle[epsfig]{article}
\font\abstract=cmr8

\font\caption=cmr8
\font\references=cmr8
\font\text=cmr10
\font\affiliation=cmssi10
\font\author=cmss10

\font\title=cmssbx10 scaled\magstep2

\null
\begin{document}
\ni
\title{Galactic Positron Production from Supernovae}

\bsk \bsk
\ni
\author{P.A.~Milne$^{1}$, M.D.~Leising$^{2}$, L.-S.~The$^{2}$}

\bsk
\affiliation{1}{ NRC/NRL Resident Research Associate,
Naval Research Lab, 
Code 7650,
Washington DC 20375}

\affiliation{2}{ Clemson University, Clemson, SC 29631}

\bsk
\baselineskip = 12pt

%\begin{abstract}
\abstract{ABSTRACT \ni
The energy deposition into the ejecta of type Ia supernovae 
is dominated at late times by the slowing of positrons 
produced in the $\beta^{+}$ decays of $^{56}$Co. Fits of 
model-generated light curves to observations of type Ia 
supernovae suggest that a significant number of positrons 
escape the ejecta and annihilate as a delayed emission. 
In this work, the isotopic yields of the $\beta^{+}$ decay 
unstable nuclei, $^{56}$Co, $^{44}$Ti \& $^{26}$Al are 
combined with the  
delayed annihilation fractions of each isotope for types 
Ia, Ib, II supernovae to generate an estimate of the 
positron production rate due to supernovae. 
It is shown that SN produced 
positrons can explain a sizeable fraction of the galactic 
positron annihilation radiation, as measured at 511 keV by the 
CGRO/OSSE, SMM and TGRS instruments.
}

\bsk
\baselineskip = 12pt

\text{\ni 1. INTRODUCTION}
\ssk
\ni
      
Supernovae (SNe), both thermonuclear and core collapse, have been suggested 
to be sources of galactic positrons through the $\beta^{+}$ decays of 
$^{56}$Co, $^{44}$Sc \& $^{26}$Al (Clayton 1973). The annihilation of 
these positrons can occur promptly (defined here as within a decade of 
the SN event) or can be delayed (defined here as $\sim$10$^{3}$ years or more 
after the SN event). The annihilation of interest in this paper is the 
delayed emission where the integrated contributions from many SNe 
combine to produce a diffuse emission. The three isotopes, 
$^{56}$Co, $^{44}$Sc \& $^{26}$Al, have very different mean lifetimes  
and thus have different factors that determine what fraction 
of the total decay positrons annihilate as a delayed emission. 
In this paper we define the parameters relevant to quantifying the 
SN contribution to galactic positrons. We then estimate these parameters, 
comparing our values with previous estimates, and emphasizing which 
estimates have observational support. In particular, we show fits to the 
late light curves of type Ia SNe which suggest that a significant 
fraction of $^{56}$Co positrons escape the SN ejecta to contribute a 
sizeable portion of the observed annihilation radiation. 

Implicit in this entire study is the assumption that all SNe are 
approximated by the types Ia, Ib \& II and that each type is 
homogeneous to the extent that a handful of models are adequate 
to describe the positron yield of the entire class. 
It is further assumed that the explosion kinematics and
nucleosynthesis of current SN modeling is accurate enough that 
parameter ranges shown truly span the range of solutions. The 
goal of this paper is to estimate the SN contribution to galactic 
positron annihilation. To perform this estimate, 
SN rates used in this study are infered from other galaxies. 
It is possible that the Galaxy has anomalous SN rates 
(as would occur if there has been a starburst at the galactic 
center within the last 10$^{6}$ years) or that there is 
leakage of positrons out of the Galaxy. Future 511 keV 
and 1.8 MeV maps (as well as maps of other nuclear decay lines) 
will be used invert this problem and {\it determine} the recent 
galactic SN history. The current study is a consistency check as 
much as it is a solution to galactic positron production. 

\bsk
\text{\ni 2. POSITRON YIELD PARAMETERS}
\ssk
\ni

The positron production rate per isotope, per SN type, is the product 
of the isotopic yield times the $\beta^{+}$ decay branching ratio, times 
the fraction of decay positrons which annihilate as a delayed emission,  
times the SN rate of that SN type. This product is then summed by 
isotope and by SN type to arrive at the total positron production rate.
 Assuming that the only relevant isotopes are 
$^{56}$Co, $^{44}$Sc \& $^{26}$Al, and that three SN types must be 
considered (types Ia, Ib \& II), quantifying the SN positron production 
rate reduces to 9 isotopic yields, 9 delayed annihilation fractions, 
3 branching ratios \& 3 SN rates. The branching ratios and other 
$\beta^{+}$ decay parameters for the three decays are relatively well-known 
and are shown in Table 1. 
The most complete estimates of the other 21 parameters was performed in 
a paper by Chan \& Lingenfelter (1993) (hereafter CL). They simulated 
positron transport through SN models of all three types arriving at a 
number of conclusions. (1) For the $\beta^{+}$ decay of $^{26}$Al, the 
mean lifetime is long-enough that $^{26}$Al is a delayed emission source 
independent of positron transport physics. For $^{26}$Al, the delayed 
fraction can be set equal to 1 for all three SN types. 
(2) The independent observation of 1.8 MeV line emission due to
the de-excitation of $^{26}$Mg$^{*}$ produced in the
$^{26}$Al $\rightarrow$ $^{26}$Mg$^{*}$ decay
can estimate the $^{26}$Al contribution (though not specifically the
SN-produced $^{26}$Al contribution) to galactic positron annihilation.
(3) For $^{56}$Co \& $^{44}$Sc decay positrons, 
the alignment and/or strength of the magnetic 
field must be considered when estimating the delayed annihilation 
fraction. For $^{44}$Sc, the effect of the magnetic field
 varies from a $\leq$1\% effect (type Ia), 
to a 66\% effect (type II). For $^{56}$Co in type Ia SNe, the 
delayed annihilation fraction changes even more dramatically. 
(4) If the magnetic field characteristics are favorable to 
positron escape, then the decay of $^{56}$Co produced in type Ia SNe 
may be the dominant contributor of galactic positrons. If not, then 
the dominant contributor of galactic positrons is from $^{44}$Sc decays.

\begin{table}
\begin{center}
\caption{{\bf TABLE 1.} Parameters relevant to the 
positron-producing decays of $^{56}$Co, $^{44}$Sc \& $^{26}$Al. }
\begin{tabular}{clccc}
\\
\hline 
\hline
\vspace{0mm}
Decay Chain & Mean Lifetime & Branching ratio & 
$<$ KE$_{e+}$$>$ (MeV)$^{a}$
& KE$^{max}_{e+}$ (MeV)$^{b}$
\\
\hline 
\\
$^{56}$Ni $\rightarrow$ $^{56}$Co $\rightarrow$ $^{56}$Fe & 
8.8$^{d}$, 111$^{d}$ & 0.19 & 0.64 & 1.46 \\
$^{44}$Ti $\rightarrow$ $^{44}$Sc $\rightarrow$ $^{44}$Ca & 
85.4$^{y}$, 6$^{h}$ & 0.95 & 0.63 & 1.46 \\
$^{26}$Al $\rightarrow$ $^{26}$Mg & 1.0 x (10$^{6}$)$^{y}$ & 0.82 & 
0.47 & 1.14 \\
\\
\hline
\hline 
%\vspace{0mm}
\multicolumn{5}{l}{$^{a}$Mean kinetic energy of distribution of
emitted positrons.} \\
\multicolumn{5}{l}{$^{b}$Endpoint energy of
distribution of emitted positrons.} \\
\end{tabular}
\end{center}
\end{table}

The critical parameter is thus the delayed annihilation fraction for 
$^{56}$Co decays in type Ia SNe, which depends upon the magnetic 
field characteristics of the SN ejecta. CL assumed that the field 
strength is strong enough to confine positrons to the field lines, 
and estimated the positron survival for two opposite geometries. 
The first geometry assumes that the field is frozen into the ejecta 
and is combed to become essentially radial as the ejecta expands. 
Positrons spiral along the field lines experiencing mirroring/beaming, 
with a fraction of the positrons escaping the SN ejecta. 
 The escaping positrons then enter a lower density 
medium and annihilate on longer timescales. This
scenario will be refered to as 
the ``radial'' scenario ($l$=$\infty$ in CL terminology).
The second geometry assumes that the field remains turbulent 
throughout the expansion of the SN ejecta. Positrons are 
trapped in the same location (in mass coordinates) as they are 
emitted. This scenario is refered to as the ``trapping'' scenario 
($l$=0 in CL terminology), and the fraction of positrons that 
survive 10$^{3}$ years in the dense ejecta is much lower than the 
escape fraction in the radial scenario. Colgate et al. (1980) argued 
for a third, ``weak field'' scenario, where the field strength is 
 inadequate 
to confine the positrons. In this scenario the positrons follow 
photon-like, straight line trajectories. Simulations have shown that 
the escape fractions and energy deposition rates for this 
scenario are approximately equal to the radial scenario (a result 
suggested by Colgate et al.). 
Throughout the remainder of this paper, only the 
terms radial and trapping will be used, but it is implied that 
radial represents radial or weak field solutions. 

\bsk
\text{\ni 3. SN Ia LIGHT CURVES \& POSITRON ESCAPE}
\ssk
\ni

CL calculated escape fractions from type Ia SNe for radial and 
trapping field geometries, but did not attempt to determine 
which scenario occurs in nature. They refered to the opposing 
conclusions of Colgate et al. (1980) and Axelrod (1980). Colgate et 
al. showed that energy deposition rates featuring 
positron escape could suitably explain the B band 
light curves of SN 1937C \& SN 1972E. 
Axelrod argued for an alternative explanation, 
that no positron escape was evidenced, rather that the apparent 
deficit in the luminosity (relative to 100\% positron trapping) 
was due to emission of an increased 
fraction of the deposited energy in unobservable infrared energy bands. 
This phenomenon was refered to as an ``infrared catastrophe'' and has 
since been observed in the core collapse supernova, 
SN 1987A. Recent papers have 
been only somewhat successful in 
clarifying the picture. Cappellaro et al. (1997) and Ruiz-Lapuente 
et al. (1997) fit model-generated bolometric light curves to 
observations, both studies concluding that positron escape is 
favored in some, but not all cases. Fransson et al. (1996), after 
including a network of reaction rates, simulated
multi-band light curves for the SN Ia model, DD4 (assuming positron 
trapping). They generated light 
curves which featured an ``infrared catastrophe'', 
but which are in conflict with the 
light curves of SN 1972E (and all other late observations of SN Ia).
None of these papers performed a detailed treatment of the positron 
transport, adapting photon transport codes instead. 

%\begin{figure}
%\centerline{\psfig{file=bol.ps,width=5in}}
%\caption{{\bf FIGURE 1.} 
%A model-generated bolometeric light curve of the model 
%W7. The
%dashed line (D) assumes instantaneous deposition of all decay energy. The
%dotted line (G) uses the results of the gamma energy deposition
%only and assumes
%no deposition of positron energy. Between these two boundaries are the
%results of the gamma energy deposition coupled with instantaneous
%positron deposition (thick line, In) and the range of curves for a  radial
%field geometry (dark shading, R) and for a trapping geometry (light
%shading, T) as the electron ionization
%fraction varies from 0.01 $\leq \chi_{e} \leq$
%3.}
%\end{figure}

Milne et al. (1999) explicitly treated both the gamma-photon 
and positron transport, assuming 
radial and trapping field geometries, and allowing for a 
range of ionizations (ranging from 1\% ionization to triple ionization). 
A characteristic model-generated bolometric light curve of the model 
W7 is shown in Figure 1. Before $\sim$200$^{d}$, positron lifetimes 
are short for all field and ionization assumptions, the light curves 
are identical with the zero lifetime (or instantaneous) light curve (In). 
With time, the SN ejecta expands and rarefies, increasing the 
positron lifetimes. In the radial scenario (R), the 
expansion leads to positron escape, and thus a deficit of energy 
deposition. As free electrons are more efficient at slowing positrons 
than are bound electrons, the low 1\% ionization light curve features the 
maximum deficit from instantaneous annihilation. In the trapping  
scenario (T), the expansion permits non-zero lifetimes, but the 
majority of the energy is later deposited. At late times, the 
deposition of this stored energy makes the trapping light curves 
brighter than either the radial or the instantaneous light curves.   
For low 1\% ionization, longer positron lifetimes lead to 
more energy being stored. The delayed deposition of this stored 
energy leads to brighter light curves at later times for low ionization 
solutions. 

\begin{figure}
\vspace{-5mm}
\centerline{\psfig{file=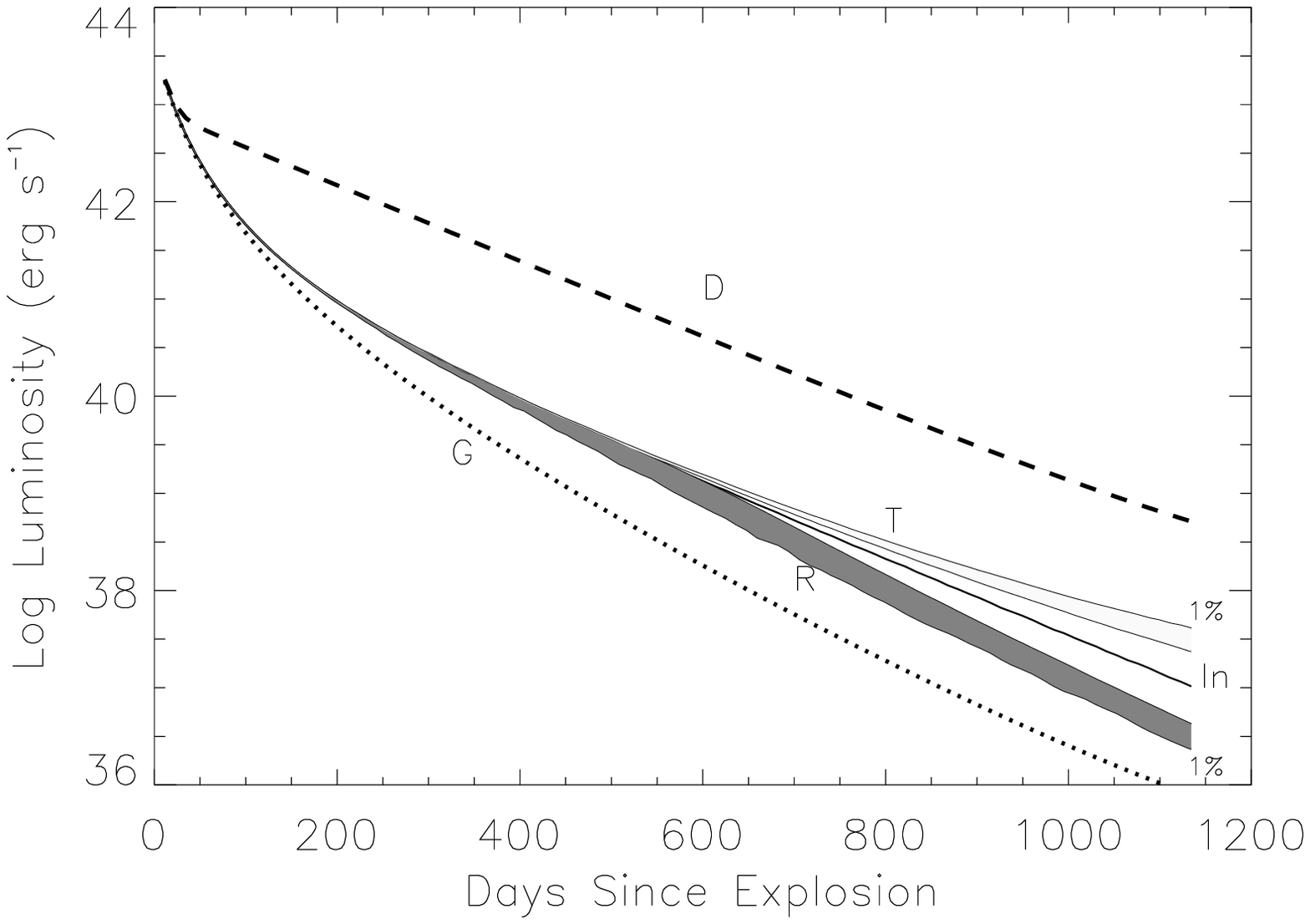,width=4.4in}}
\vspace{-3mm}
\caption{{\bf FIGURE 1.}
A model-generated bolomteric light curve of the model
W7. The
dashed line (D) assumes instantaneous deposition of all decay energy. The
dotted line (G) uses the results of the gamma energy deposition
only and assumes
no deposition of positron energy. Between these two boundaries are the
results of the gamma energy deposition coupled with instantaneous
positron deposition (thick line, In) and the range of curves for a  radial
field geometry (dark shading, R) and for a trapping geometry (light
shading, T) as the electron ionization
fraction varies from 0.01 $\leq \chi_{e} \leq$
3.\vspace{5mm}}
\makebox[3in][l]{\psfig{figure=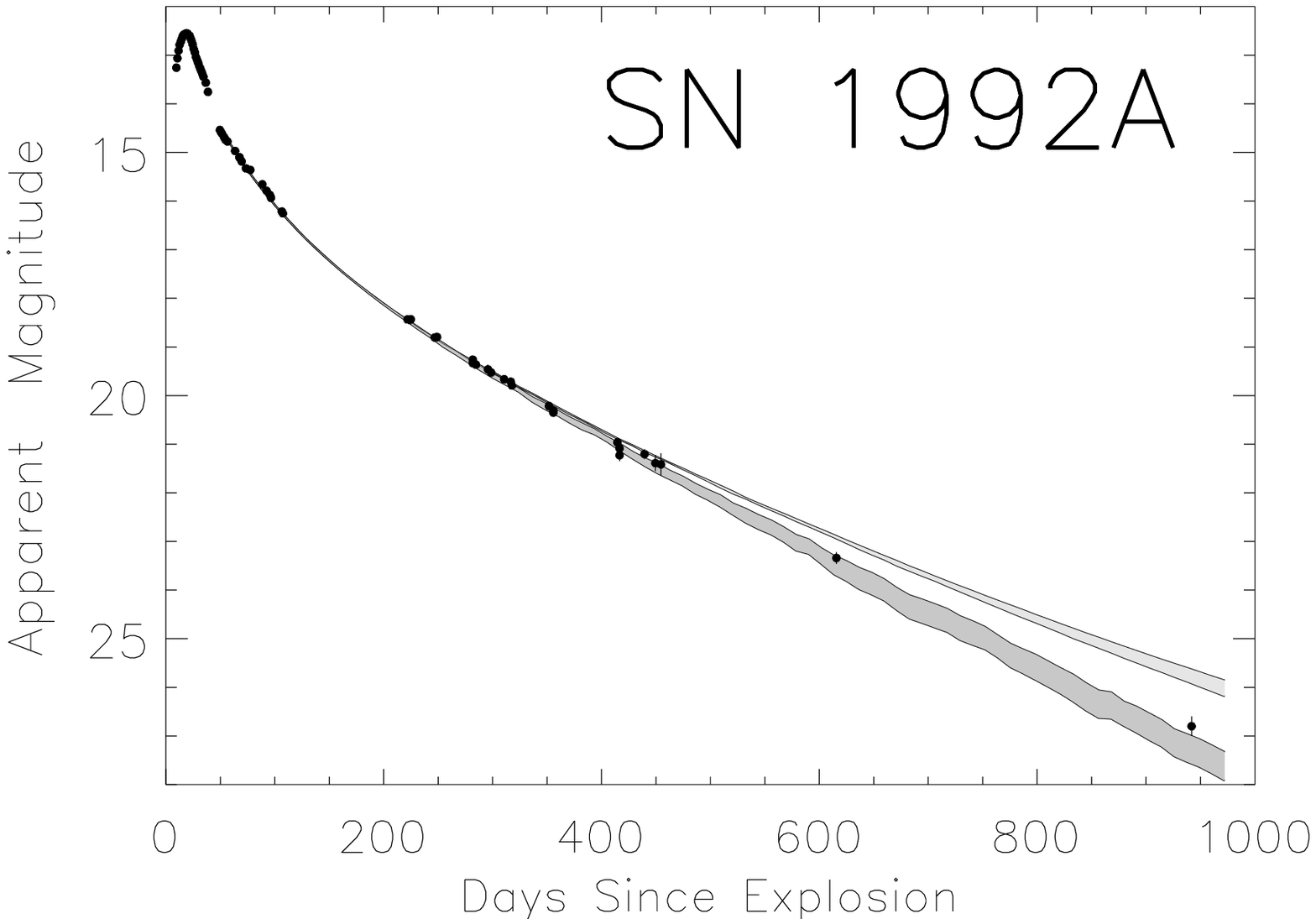,width=3in}}
\makebox[3in][l]{\psfig{figure=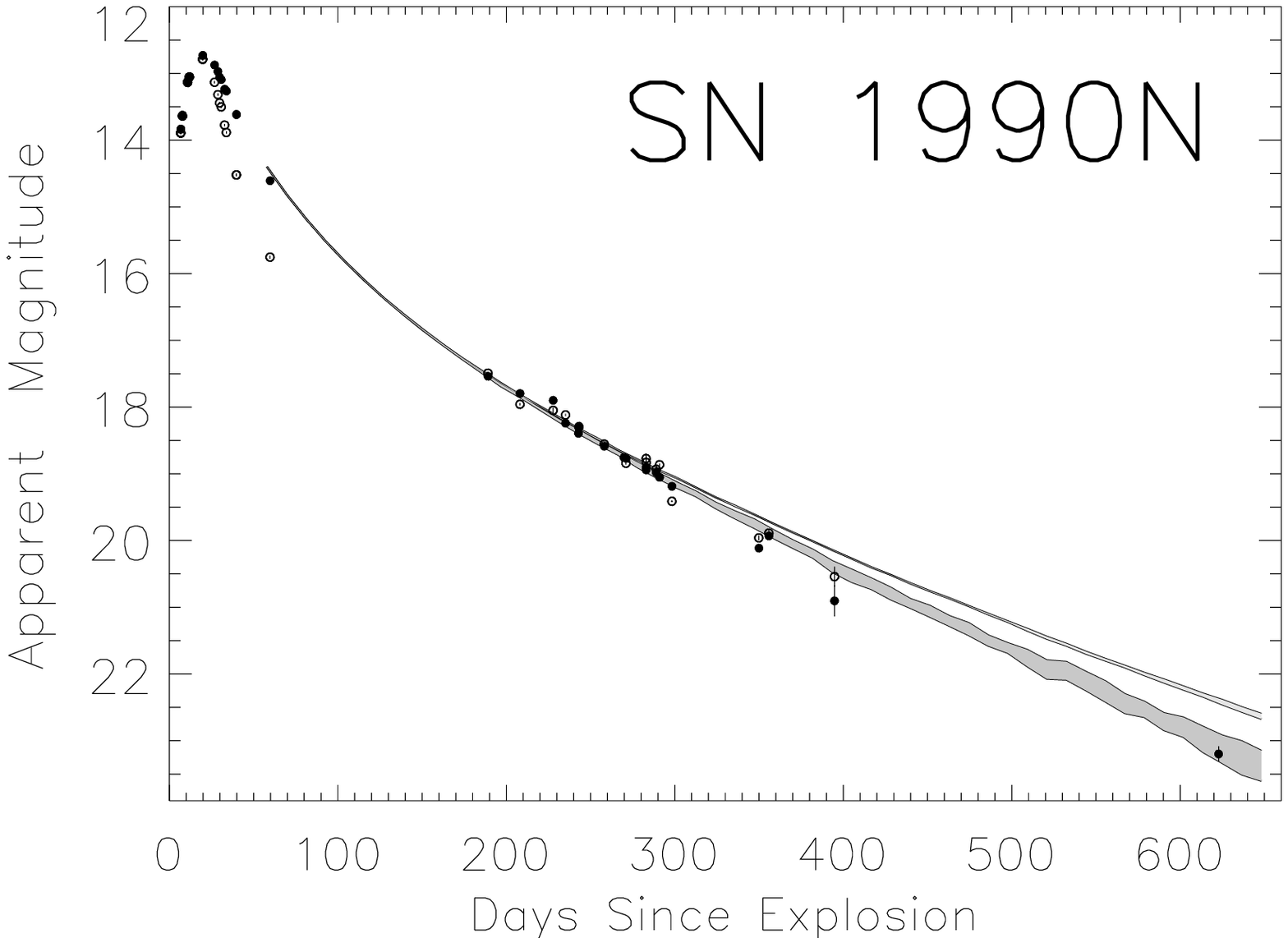,width=3in}}
\linebreak
\makebox[3in][l]{\psfig{figure=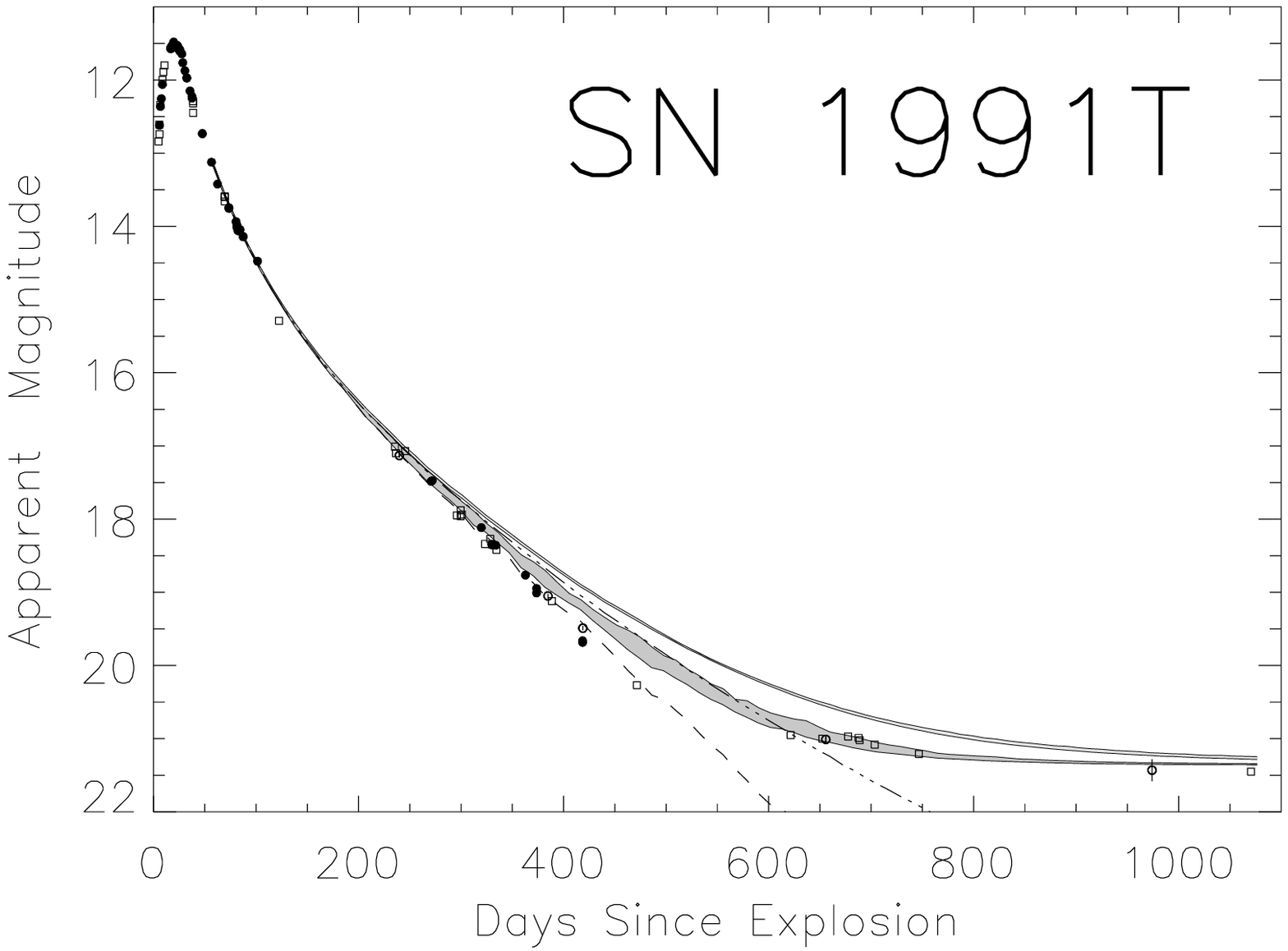,width=3in}}
\makebox[3in][l]{\psfig{figure=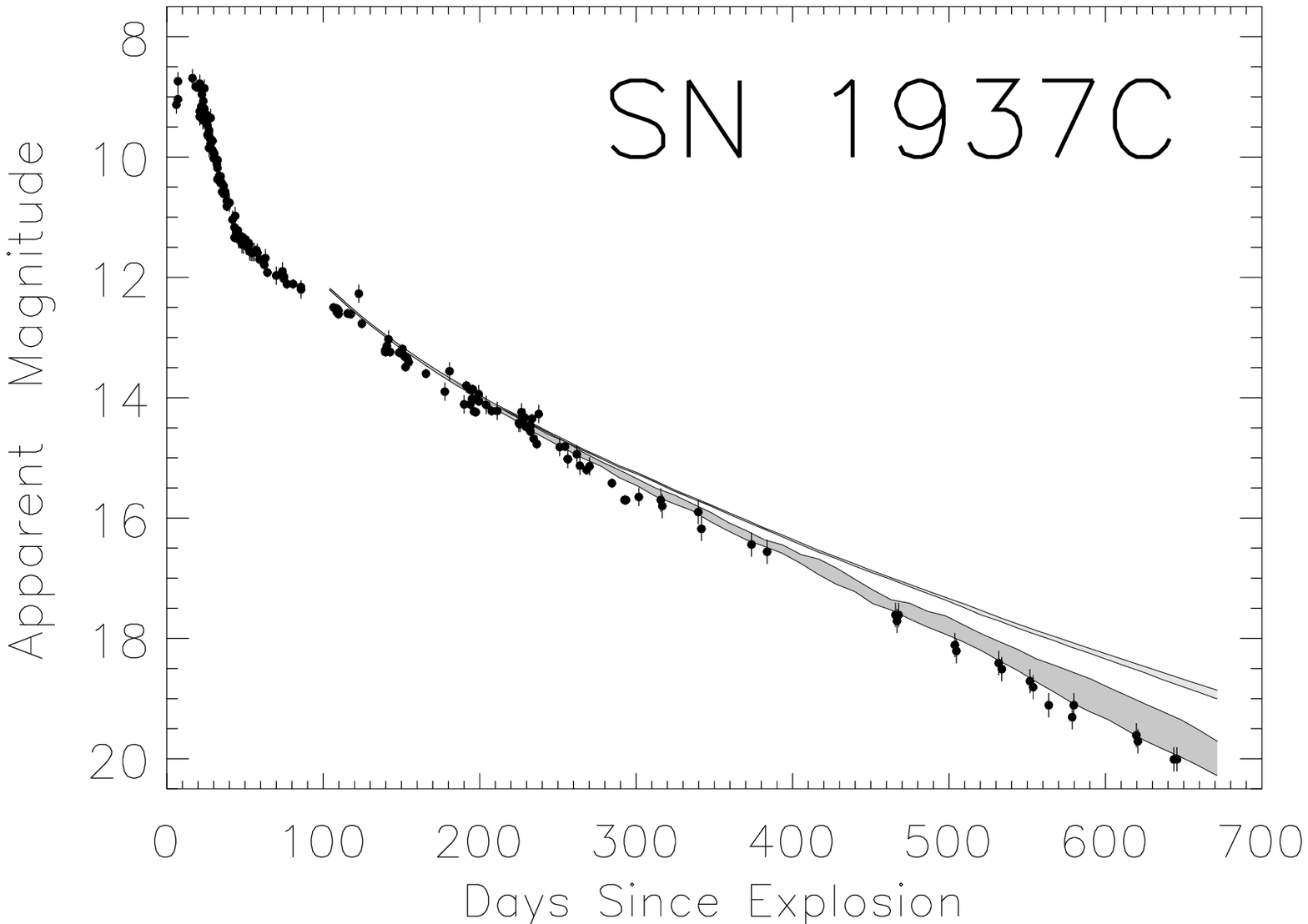,width=3in}}
\linebreak
\caption{{\bf FIGURE 2.} Model-generated light curves fit
to the V and B band light curves of
four type Ia SNe; SN 1992A (V: Suntzeff 1996), SN 1990N (B\&V:
Lira et al. 1998),
SN 1991T (V: Lira et al. 1998, Cappellaro et al. 1997, Schmidt et al. 1994)
 \& SN 1937C (B: Schaefer 1994). The light curves are better explained
by radial escape (dark shading) than by positron trapping (light
shading). The model-generated light curves of SN 199T are shown
with a constant luminosity light echo added, the same curves without
a light echo are displayed with dashed (radial) and dot-dashed
(trapping) lines.}
\end{figure}

Bolometric light curves were generated for 22 SN Ia models. These 
models were selected to span  
Chandrasekhar, sub-Chandrasekhar and merger scenarios. The separation 
between radial and trapping light curves was found to exceed either 
the effects of model type or the level of ionization. 
These model-generated light curves were then fitted to the 10 SN Ia 
best observed at late times.\footnote{Discussions regarding the 
validity of fitting 
model-generated energy deposition rates to observed band photometry 
and/or ``uvoir'' bolometric light curves are given in Milne et al.
(1999).} Of these 10 SNe, eight were considered 
either normal- or super-luminous. Of these 8, seven were suitably 
fitted with positron escape according to the radial field 
scenario. Five of the seven were better fitted with positron 
escape than with positron trapping. None were better fitted with 
positron trapping. The four best examples are shown in Figure 2, 
all 10 SNe are shown in Milne et al. (1999). The preference of 
positron escape is apparent in SNe 1992A (both bolometric (not shown) and 
V band, fit with DD23C (H\H{o}flich et al. 1998)) \& 1990N 
(fit with W7DN (Yamaoka et al. 1992)) without 
further explanation. For SN 1937C (fit with DET2 (H\H{o}flich, 
Khokhlov, \& Wheeler 1995)), B band data was used, thus 
fitting began after 120$^{d}$ at which time color evolution has been 
empirically determined to cease in SN Ia. For SN 1991T 
(fit with HECD (Kumagai 1997)), the very late 
emission is dominated by a light echo, discovered spectrally by 
Schmidt et al. (1996) and imaged by HST (Boffi et al. 1998). Though 
the light curves for the range of ionizations (filled curves) 
are shown with a 
constant luminosity light echo included, the dashed and dot-dashed 
lines show the low ionization radial and trapping curves with no 
light echo. It is apparent that the trapping curves are already too 
bright without the additional emission from an echo, the radial 
curve suitably explains the data before 480 days. This suggests a 
light echo that ``turns on'' after 480 days, which would occur if the 
light echo is due to the peak light sweeping through an off-axis cloud.
The asymmetry of the HST image would appear to permit that interpretation.

The conclusion of Milne et al. (1999) is that 
the late light curves of normal- and super-luminous 
type Ia SNe can be suitably explained with positron escape. Though 
the sample is neither large enough, nor well-enough observed to 
rule out trapping in all cases, the lack of counter-examples suggest 
escape to be the dominant result. Many of 
the SNe used in that study were also used in the studies of 
Cappellaro et al. (1997), Ruiz-Lapuente et al. (1997) \& Fransson 
et al. (1996). Milne et al. (1999) did not find positron trapping to 
be favored for any SN, in mild disagreement with some of the 
conclusions of these earlier studies. Based upon these observations, 
the positron escape fractions claimed in Milne et al. (1999) will be 
inserted into the $^{56}$Co-SN Ia portion of the positron 
production rate calculations rather than the trapping-radial 
range suggested by CL. The determination of the escape of 
$^{56}$Co positrons from type Ia SNe was based upon the existence of 
a ``positron phase'', an epoch during which the energy deposition is  
dominated by the slowing of positrons. This ``positron phase'' 
does not occur in core collapse SNe, due to the more efficient 
trapping of gamma-photons. The dominance of $^{56}$Co decay 
photons transitions to the dominance of $^{57}$Co decay photons 
without an epoch of positron dominance. 
Since no determination of the field characteristics of types Ib 
and II SNe are possible, the trapping-radial ranges claimed by CL 
are used in section 5.   

\bsk
\text{\ni 4. ISOTOPIC YIELDS}
\ssk
\ni

The SN Ia delayed annihilation fraction may prove to be the most 
critical parameter for quantifying positron production from SNe, 
but twenty other parameters must also be estimated. The other eight 
delayed annihilation fractions were adequately approximated by CL, 
the nine isotopic yields and three SN rates remain undetermined. 

The isotopic yields of $^{56}$Ni (the parent isotope of $^{56}$Co) 
are the best constrained. In all three SN types, energy deposition 
from the $^{56}$Ni and subsequent $^{56}$Co decays power the light 
curves. This allows fits of model-generated bolometric light curves 
to observations, 
combined with distance estimates to the host galaxy, to constrain 
the $^{56}$Ni yields. Type Ia SNe produce the most $^{56}$Ni per 
SN event, ranging from 0.3 -0.9 M$_{\odot}$ in various SN Ia models. 
Diehl (1997) suggests the typical $^{56}$Ni yield to be 0.5 M$_{\odot}$. 
As shown by Milne et al. (1999), the larger nickel yield in 
Chandrasekhar mass models partially compensates for the lower 
escape fraction (relative to sub-Chandrasekhar mass models), leading 
to the positron yields being virtually independent of the 
progenitor mass. The SN Ia $^{56}$Ni yield estimate is 
also influenced by the existence of super- and sub-luminous classes. An 
additional result of Milne et al. (1999) is the acceptance of 
both Chandrasekhar mass and sub-Chandrasekhar mass explanations 
for normally- and super-luminous SNe Ia, but rejection of all 
currently proposed models for sub-luminous SNe Ia. CL discussed a 
``Ip'' model class as an explanation for sub-luminous SNe Ia. The 
Ip model class featured large positron yields. This work 
will not include that class due to the findings of Milne et al. (1999).

The $^{56}$Ni yields from type II SNe are well-constrained due to 
the observations of the nearby SN 1987A. That SN produced 0.08 M$_{\odot}$ 
of $^{56}$Ni, the fiducial value used by Dermer \& Skibo (1997). 
The type Ib $^{56}$Ni yield is taken from CL, ranging from 0.08 -0.28 
M$_{\odot}$. 
Diehl (1997) estimates the type II/Ib $^{56}$Ni yield to be 0.1 M$_{\odot}$. 
The type II/Ib $^{56}$Ni yields are of little importance due to the low 
delayed annihilation fractions for these SNe. 

The isotopic yields of $^{44}$Ti are obtained from the outputs of SN 
models (with fewer observational constraints) and are poorly constrained. 
Timmes et al. (1996) show a range of $^{44}$Ti yields for core 
collapse models, claiming 3~x~10$^{-5}$ M$_{\odot}$ (type II) \& 
6~x~10$^{-5}$ M$_{\odot}$ (type Ib) to be typical values, but later 
tripling these yields in a positron production calculation. 
Nomoto
(1997) suggests that type II SNe produce 5~x~10$^{-5}$ M$_{\odot}$
of $^{44}$Ti, in agreement with the value used by Diehl (1997) for 
types II/Ib. Meyer et al. (1995) estimates the $^{44}$Ti yield 
from a 25 M$_{\odot}$ SN to be 1.5~x~10$^{-4}$ M$_{\odot}$. 
The $^{44}$Ti $\rightarrow$ $^{44}$Sc $\rightarrow$ $^{44}$Ca 
decay has been directly observed in the Cas A SNR (type II or Ib) 
via the 1.157 MeV de-excitation 
line of $^{44}$Ca and the 68 keV and 78 keV de-excitation lines of 
$^{44}$Sc. From measurements taken by the CGRO/COMPTEL, the 
CGRO/OSSE and the {\it Rossi X-Ray Timing Explorer} High Energy 
X-Ray Timing Experiment, the 
$^{44}$Ti yield is estimated to be 2.2~x~10$^{-4}$ 
M$_{\odot}$ (The et al. 1998, using the 85$^{y}$ $^{44}$Ti 
mean lifetime from Ahmad et al. 1997).
The SN Ia $^{44}$Ti yields are equally uncertain. 
The Chandrasekhar mass 
model, W7 (Nomoto, Theilemann \& Yokoi 1984), produces $\sim$1.8 x 
10$^{-5}$ M$_{\odot}$ of $^{44}$Ti. The low mass portion of 
 sub-Chandrasekhar mass models 
produce more $^{44}$Ti, ranging from (2-40)~x~10$^{-4}$ M$_{\odot}$ 
(Woosley \& Weaver 1994), though these low mass examples cannot 
explain 100\% of SN Ia events due to nucleosynthesis considerations. 
Dermer \& Skibo (1997) use the CL results, 
scaling the $^{44}$Ti yield with the $^{56}$Ni yields. This leads to 
(3-8)~x~10$^{-5}$ M$_{\odot}$, 1.7~x~10$^{-4}$ M$_{\odot}$ \& 
$\leq$ 2.0~x~10$^{-4}$ M$_{\odot}$ per SN for types Ia, Ib, II respectively.  

The isotopic yield of $^{26}$Al in type Ia SNe is very low 
($\sim$ 10$^{-6}$ -Diehl 1997). In core collapse SNe, estimates 
of the yields range from 2~x~10$^{-4}$ M$_{\odot}$ (for SN II/Ib 
-Diehl 1997) to (0.3 -20)~x~10$^{-5}$ M$_{\odot}$ (for SN II 
-Prantzos 1996). Timmes \& Woosley (1997) suggest the mass function 
averaged $^{26}$Al yield per core collapse SN event is 
7.7~x~10$^{-5}$ M$_{\odot}$. 

\bsk
\text{\ni 4. SUPERNOVA RATES}
\ssk
\ni

Estimation of SN rates in the Galaxy is dependent upon the 
collective SN rates from other spiral 
galaxies.\footnote{Estimates of the galactic SN rate from 
the historical record, from surveys of supernova remnants, 
and from galactic nucleosynthesis are considered to 
be less reliable than assuming the Galaxy to be an Sb galaxy 
and assuming that the SN rates scale with L$^{B}$.} The SN rates 
from SN surveys are given in units of 10$^{10}$ L$_{\odot}^{B}$ 
and scaled to 2.3~x~10$^{10}$ L$_{\odot}^{B}$. The differences 
between suggested rates is due to different algorithms 
accounting for selection effects. The SN Ia rate is relatively 
well-constrained, estimates are 0.4$\pm$0.1 (Cappellaro et al. 
1997), 0.28 (Tammann et al. 1994), 0.5 (Hatano et al. 1997). 
The SN Ib rate is approximately equal to the SN Ia rate, but 
the estimates are more varied. Estimates are 0.2$\pm$0.1 
(Cappellaro et al. 1997), 0.8 (Hatano et al. 1997), 
0.3 (Tammann et al. 1994). The SN II rate is the largest of the 
three and the most varied. Estimates are 1.2$\pm$0.6 
(Cappellaro et al. 1997), 3.8 (Hatano et al. 1997), 
1.5 (Tammann et al. 1994). 

\bsk
\text{\ni 5. POSITRON PRODUCTION RATES}
\ssk
\ni

Shown in Table 2 are the parameters relevant to estimating the 
galactic positron production rate from SNe. In many cases, the 
most favorable positron yields per SN event, as shown in 
column (6), are realized only by anomalous SNe, and would 
have large recurrence times. As seen in columns 
(6) and (7), SN Ia $^{56}$Co decays yield the most positrons 
per SN and per second. The $^{56}$Co decay contributions from SN 
Ib \& II are zero due to rejecting the 100\% mixing scenario 
allowed by CL. The $^{44}$Ti decay contributions (per second) 
are not negligible. The favorable $^{44}$Ti yields cannot be 
claimed for all three SN types simultaneously, as that would 
lead to anomalously large $^{44}$Ca/$^{56}$Fe production. The 
largest $^{44}$Ti decay contribution would result if a 
considerable fraction of all SN Ia events are low-mass 
sub-Chandrasekhar explosions, and core collapse SNe possess 
radial (or weak) magnetic fields. The $^{26}$Al decay 
contribution is dominated by core collapse SNe. The 
positron yield per second is better estimated by including 
results of measurements of the 1.8 MeV line emission. 
In 82\% of $^{26}$Al $\rightarrow$
$^{26}$Mg decays, both a positron and a 1.8 MeV line photon are
produced. Core collapse SNe have been suggested to
account for as much as 100\% of galactic $^{26}$Al (Timmes et al. 
1997). Assuming the entire emission to emanate from core collapse 
SNe at the distance of the   
galactic center ($\sim$ 8 kpc distant), 
reported 1.8 MeV central radian fluxes of 
3~x~10$^{-4}$ ph cm$^{-2}$ s$^{-1}$ (Knodlseder et al. 1999) 
translates to 1.9~x~10$^{42}$ e+ s$^{-1}$, 
approximately equal to the $^{44}$Ti decay production rate. 

%Assuming a constant SN rate over the last 10$^{6}$ years, steady-state 
%positron production/annihilation and no leakage of positrons from the 
%Galaxy, the above production rate can be compared with the 
%511 keV line flux. A positron production rate of 1.4 x 10$^{43}$ e+ s$^{-1}$ 
%would generate a 511 keV flux of 1.1 x 10$^{-3}$ ph cm$^{-2}$ s$^{-1}$ 
%if emitted from the galactic center ($\sim$ 8 kpc distant) and 
%annihilating with a positronium fraction of 0.95. That is about 
%1/3 - 1/2 of the total 511 keV flux suggested by CGRO/OSSE, SMM \& 
%TGRS measurements (Milne et al. 1999). Correcting this value to a 
%distributed emission has not been performed.\footnote{Timmes et al. (1997) 
%quoted a positron production rate of 1.5 x 10$^{-3}$ e+ s$^{-1}$ from 
%511 keV measurements, that value is too low by a considerable fraction.}

\begin{table}
\begin{center}
\caption{{\bf TABLE 2.} Galactic positron production rates from 
SNe. Italicized values denote prefered values from a range.}
\begin{tabular}{lllcccc}
\\
\hline
\hline
SN & SN & Isotope & Isotopic & Delayed & 
\multicolumn{2}{c}{Positron Yield} \\
Type & Rate$^{(a)}$ & & Yield (M$_{\odot}$) & Ann. Frac.& 
[per SN]$^{(b)}$ & [per s]$^{(c)}$ \\
(1) & (2) & (3) & (4) & (5) & (6) & (7) \\
\hline 
\\
Ia & 0.4$\pm$0.1 & & & & & \\
&   & $^{56}$Ni & {\it 0.5}(0.2-0.8) & 0.001-0.11 & 
{\it 8}(1-20)$^{(d)}$ & {\it 1.0}(0.8-1.3) \\
 &  & $^{44}$Ti & ({\it 0.2}-12) x 10$^{-4}$ & 0.99 & 
0.04-3.8 & {\it 0.01}-0.49 \\
 &  & $^{26}$Al & 10$^{-6}$ & 1.00 & 4 x 10$^{-3}$ & 4~x~10$^{-4}$ \\
Ib & 0.3$^{+0.5}_{-0.1}$ & & & & & \\
  & & $^{56}$Ni & 0.18$\pm$0.10 & 0.0 & 0.00 & 0.00 \\
  & & $^{44}$Ti & ({\it 6}-22)~x~10$^{-5}$ & 
0.85-{\it 0.99} & 0.1-0.4 &{\it 0.01}-0.29 \\
  & & $^{26}$Al & {\it 0.8}(0.3-20)~x~10$^{-4}$ & 1.00 & 
0.01-0.91 & {\it 0.03}(0.00-0.23) \\
II & 3.1$^{+0.7}_{-1.9}$ & & & & & \\
  & & $^{56}$Ni & 0.08-0.28 & 0.0 & 0.00 & 0.00 \\
  & & $^{44}$Ti & ({\it 3}-22)~x~10$^{-5}$ & 0.34-{\it 0.99} & 
0.03-0.62& {\it 0.21}(0.01-0.79) \\
  & & $^{26}$Al & {\it 0.8}(0.3-20)~x~10$^{-4}$ & 1.00 & 
0.01-0.91 &{\it 0.30}(0.00-1.11) \\
\\
\hline
\hline
\multicolumn{7}{l}{$^{(a)}$ The SN rate is in units of SNe per century.} \\
\multicolumn{7}{l}{$^{(b)}$ The yields per SN are in units of 10$^{52}$ 
positrons SN$^{-1}$.} \\
\multicolumn{7}{l}{$^{(c)}$ The yields per second are in units of 10$^{43}$
positrons s$^{-1}$.} \\
\multicolumn{7}{l}{$^{(d)}$ The SN Ia yield per SN is the from Milne et al. 
(1999), rather than the product of SN Ia parameters.} \\
\end{tabular}
\end{center}
\end{table}

The total positron production rates are 1.0~x~10$^{43}$ e+ s$^{-1}$ for
$^{56}$Co decays, 2.4~x~10$^{42}$ e+ s$^{-1}$ for $^{44}$Ti decays, and
1.9~x~10$^{42}$ e+ s$^{-1}$ for $^{26}$Al decays. These values are in
general agreement with estimates by Timmes et al. (1996), who derived
the values 1.6~x~10$^{43}$ e+ s$^{-1}$, 4.5~x~10$^{42}$ e+ s$^{-1}$,
and 2.9~x~10$^{42}$ e+ s$^{-1}$ respectively.
Notably different is the $^{44}$Ti yield. In that study, it was 
erroneously suggested that two positrons are generated in 95\% of 
$^{44}$Ti $\rightarrow$ $^{44}$Sc $\rightarrow$ $^{44}$Ca decays 
(a single positron is generated). 
That study also tripled the $^{44}$Ti
yields in all three SN types to permit their chemical evolution
model to match the solar $^{44}$Ca abundance, but used the lower 
delayed annihilation fraction (34\% rather than the value 99\% used 
in this study). The net difference is that the $^{44}$Ti yields 
in Timmes et al. (1996) are roughly double the values claimed here.
 The difference in
$^{26}$Al decay yield may be due to a different 1.8 MeV total flux.

Assuming a constant SN rate over the last 10$^{6}$ years, steady-state
positron production/annihilation and no leakage of positrons from the
Galaxy, the above production rate can be compared with the
511 keV line flux. A positron production rate of 1.4~x~10$^{43}$ e+ s$^{-1}$
would generate a 511 keV flux of 1.1~x~10$^{-3}$ ph cm$^{-2}$ s$^{-1}$
if emitted from the galactic center ($\sim$ 8 kpc distant) and
annihilating with a positronium fraction of 0.95. That is about
1/3 - 1/2 of the total 511 keV flux suggested by CGRO/OSSE, SMM \&
TGRS measurements (Milne et al. 1999b). Correcting this value to a
distributed emission has not been performed.\footnote{Timmes et al. (1997)
quoted a positron production rate of 1.5~x~10$^{-3}$ e+ s$^{-1}$ from
511 keV measurements, that value is too low by a considerable fraction.}

\pagebreak
\bsk
\text{\ni 6. DISCUSSION}
\ssk
\ni

The positrons produced in the $\beta^{+}$ decays of $^{56}$Co,
$^{44}$Ti \& $^{26}$Al have been shown to potentially contribute a
large fraction of the total galactic positron production rate. This
work shows observational support for the the suggestion made by
Chan \& Lingenfelter (1993) that significant numbers of
$^{56}$Co decay positrons can escape from the ejecta of type Ia SNe.
Observations of supernovae and supernova remnants, combined with
improved nucleosynthesis modeling has led to improved constraints
upon the various parameters that determine SN positron production.

A future step in modeling the galactic positron production rate from
SNe will be to compare spatial distributions with CGRO/OSSE measurements
of 511 keV positron-electron annihilation radiation. Type II/Ib SNe are
seen only in the disks of spiral galaxies, type Ia SNe occur in
both bulges and disks. More progress must be made towards understanding
both the intrinsic B/D ratios of type Ia SNe in spiral
galaxies, and the B/D ratio measured in the 511 keV emission before
useful comparisons can be made.

An observation that would
solidify the picture of SN positron production would be the detection
of 511 keV emission from nearby SNe. In type Ia SNe, detection of
emission above the level expected for $^{44}$Ti decays would be
evidence of positron escape. In all SN types, the intensity
and distribution of the emission would trace the diffusion of
positrons from the remnant into the Galaxy. Additionally, the
detection of prompt positron annihilation in SNe would better
constrain isotopic yields and delayed annihilation fractions.
The late light curves of SNe Ia suggest positron escape, but
more observations are required to accurately quantify the
positron yields. If all these observations are made, the
hypothesis of SN production of galactic positrons will be
demonstrated and positron annihilation radiation measurements
can be used to probe the recent SN history in the Galaxy.
                                                                                                                                
%\begin{references}

{\references \ni REFERENCES
\ssk

\noindent Ahmad, I. et al. 1997, Phys. Rev. Lett., 80, 2550

\noindent Axelrod, T.S. 1980, Ph.D.thesis, Univ. California at 
Santa Cruz

\noindent Boffi, F.R. et al. 1998, BAAS, 192, 6.07

\noindent Cappellaro, E., et al. 1997a, A \& A, 322, 431

\noindent Cappellaro, E., et al. 1997b, A \& A, 328, 203

\noindent Chan, K.-W., Lingenfelter, R. 1993, ApJ, 405, 614

\noindent Clayton, D.D. 1973, Nature, 244:139, 137 

\noindent Colgate, S., Petschek, A.G., Kreise, J.T. 1980, ApJ, 237, L81

\noindent Dermer, C.D., Skibo, J.G. 1997, ApJ, 487, L57

\noindent Diehl, R. 1997, in Proceedings of the 2nd INTEGRAL Workshop, 9

\noindent Fransson, C., Houck, J., Kozma, C. 1996, in IAU Colloq. 145, 
Supernovae and
Supernova Remnants, ed. R. McCray \& Z. Wang (Cambridge:
Cambridge University Press), 41
                                                       
\noindent Hatano, K., Fisher, A., Branch, D. 1997, 290, 360

\noindent H\H{o}flich, P., Khokhlov, A., Wheeler, J.C. 1995, ApJ, 444, 831

\noindent H\H{o}flich, P., Wheeler, J.C., Theilemann, F.-K. 1998, 
ApJ, 495, 617

\noindent Knodlseder, J., et al. 1999, A \& A, 344, 68

\noindent Kumagai, S. 1997, private communication

\noindent Lira, P. et al. 1998, AJ, 115, 234

\noindent Meyer, B.S., Weaver, T.A., Woosley, S.E. 1995, 
Meteoritics, 30, 325

\noindent Milne, P.A., The, L.-S., Leising, M.D. 1999, ApJS, 124, 503

\noindent Milne, P.A., et al. 1999b, these Proceedings, ,

\noindent Nomoto, K., Theilemann, F.-K., Yokoi, K. 1984, ApJ, 286, 644

\noindent Nomoto, K., et al. 1997, Nuclear Physics A, 621, 467

\noindent Prantzos, N., Diehl, R. 1996, Phys. Rep., 267, 1

\noindent Ruiz-Lapuente, P., Spruit, H. 1997, ApJ, 500, 360

\noindent Schaefer, B.E. 1994, ApJ, 426, 493

\noindent Schmidt, B.P. et al. 1994, ApJ, 434, L19

\noindent Sunzteff, N.B. 1996, in IAU Colloq. 145, Supernovae and 
Supernova Remnants, ed. R. McCray \& Z. Wang (Cambridge: 
Cambridge University Press), 41

\noindent Tammann, G.A., L\H{o}ffler, W., Schr\H{o}der, A. 1994, 
ApJS, 92, 487

\noindent The, L.-S., et al. 1998, ApJ, 504, 500

\noindent Timmes, F.X., et al. 1996, ApJ, 464, 322

\noindent Timmes, F.X., Diehl, R., Hartmann, D.H. 1997, ApJ, 479, 760 

\noindent Timmes, F.X., Woosley, S.E. 1997, ApJ, 481, L81

\noindent Yamaoka, H., et al. 1992, ApJ, 393, L55
}
\end{document}